\documentclass[11pt]{article}
\newcommand{\be}{\begin{equation}}
\newcommand{\ee}{\end{equation}}
\newcommand{\ba}{\begin{eqnarray}}
\newcommand{\ea}{\end{eqnarray}}

\begin{document}

\title{A note on the structural stability of tachyonic inflation}

\author{ \large J.M. Aguirregabiria and Ruth Lazkoz\\
{\it  \small Fisika Teorikoa eta Zientziaren Historia Saila,}\\
{\it \small Zientzia eta Teknologiaren Fakultatea, 
}\\
{\it \small  Euskal Herriko Unibertsitatea, 644 Posta Kutxatila, 48080 Bilbao, Spain}}
\date{}
\maketitle
\begin{abstract}

We study the structural stability of tachyonic inflation against changes in the shape of the potential. Following Lidsey (Gen. Rel. Grav. {\bf 25}, 399 (1993)),  the concepts of rigidity and fragility are defined through a condition on the functional form of the Hubble factor.  We find that the models are rigid in the sense that the attractor solutions never change as long as the conditions for inflation are met. 

\end{abstract}

\maketitle
The inflationary paradigm has received the support of WMAP data \cite{WMAP}, but a completely satisfactory theoretical explanation remains nonexistent. Some works in this direction have attempted to find an answer in string theory, and as a result 
tachyonic inflation has been put forward \cite{sen}. The idea strongly relies in the possibility of describing tachyon condensates in string theories
in terms of perfect fluids. Many papers have been devoted to the study of cosmological consequences of such fluids  both in general relativity and brane-world cosmology \cite{others, attractors, brane}. 

Recently, tachyonic inflation has been criticized \cite{critics}. It seems though, that the situation is not so clear cut because potentials have been found which seem to circumvent the problem \cite{brane}.

This area of cosmological modelling, as any other,  involves a great deal of idealization; any conclusion or result may turn out to depend strongly on the chosen values and number of free parameters in the tachyon potential. Simplifying assumptions made in common practice do not rely on observations, but are rather chosen for technical convenience. Besides, one   cosmological observations are model dependent  and not devoid of errors. Thus, making assumptions based on observations  results in better but not perfect modelling. 
As put forward many years ago by Andronov and Pontryagin \cite{stst}, satisfactory models of real phenomena, cosmology included, should be structurally stable, that is,  should provide predictions qualitative independent of perturbations. In addition, it has been claimed \cite{coley} that the concepts of rigidity and fragility seem to be important for most cosmological models, and that structural fragility might be the suitable theoretical setup for cosmology \cite{tavakol}.

 Turning back to  tachyonic inflation, we address the problem of its  rigidity or fragility  from the perspective of tachyon field dynamics. We follow closely an approach devised by Lidsey for standard scalar fields \cite{lidsey}. The method allows identifying regions where
the attractor solutions change, thus indicating fragility in the system (see \cite{attractors} for discussions on  tachyonic inflationary attractors).

The evolution equations for a flat ($k=0$) Friedmann-Robertson-Walker (FRW) cosmological
model with a tachyon field $T$ evolving in a potential $V(T)$  are
\begin{equation}
\frac{2\dot H}{3 H^2}=-\dot T^2\label{eq:raychaudhuri}
\end{equation}
\begin{equation}
\frac{\ddot T}{1-\dot T^2}+3H\dot T+\frac{V,_T}V=0,\label{eq:kleingordon}
\end{equation}
\begin{equation}
\dot\rho_{\gamma}+3\gamma H\rho_{\gamma}=0,\label{eq:cont}
\end{equation}
which are, in turn, subject to the Friedmann constraint
\be
3 H^2=\frac{V}{\sqrt{1-\dot T^2}}\label{friedmann}.
\ee
Here and throughout overdots denote differentiation with respect to cosmic time $t$, $H\equiv  \dot a/a $ is the 
Hubble parameter, and $a$ the synchronous scale factor.

Specifically,  we will look for correspondences between the space of scale factors and that of inflationary potentials. We will set the discussion in a general framework (valid for any potential)
which is an adaptation of Lidsey's
approach.

Using the Hamiltonian formalism the Friedmann constraint can be cast in the following form:
\be
({H'(T)})^2-\frac{9}{4}H^4(T)+\frac{1}{4}V(T)^2=0\label{hamilton},
\ee
where $H'(T)\equiv dH(T)/dT$.
Solutions to (\ref{hamilton}) can be labelled by means of a parameter $p$, so that we have $H(T(t),p)$. The value of $p$ is fixed unambiguously once the initial conditions have been
chosen. The corresponding expression for the scale factor will be
\be
a(T(t),p)=a_i \,{\rm exp}\left(-\frac{3}{2}\int_{T_i}^{T}d\tilde T H^3(\tilde T,p)\left(\frac{\partial H(\tilde T,p)}{\partial \tilde T}\right)^{-1}\right),
\ee
where $a_i$ and $T_i$ are constants of integration. 

Let us consider now two solutions $H(T,p+\Delta p)$ and $H(T,p)$, under the requirement that
they are very close together in the corresponding space, i.e. $\vert\Delta p\ll 1\vert$.  
We then have 
\be
H(T,p+\Delta p)-H(T,p)\approx\left(\partial H/\partial p\right)_T\Delta p\,.\ee
By differentiating  (\ref{hamilton}) with respect to $p$, and combining the result 
with the evolution equation (\ref{eq:raychaudhuri}) and the definition of $H$ we get
\be
H(T,p+\Delta p)-H(T,p)\propto a^{-3}(T,p)\Delta p\label{eq:ha}\,.
\ee
As a consequence, the differences between not very different solutions get washed off in the course of the evolution, which means they approach some 
attractor solution $H(T)$. However, the attractor may not be the same for all values
of the potential parameters; in other words, the system may be fragile around the point
at which the attractors change. In order to check whether that is the case, one defines
the quantity
\be
F\equiv\left\vert\frac{H(T,p+\Delta p)}{H(T,p)}-1\right\vert\,.
\ee
Equation (\ref{eq:ha}) shows that in an expanding universe $F\to0$ as time grows, but the form of the attractor
may vary if $\partial F/\partial T$ changes sign. Note that $F$ can go to zero for $\partial F/\partial T>0$ or for $\partial F/\partial T<0$, but not for both. Thus, 
if  $\partial F/\partial T=0$ for some value of the parameters the system will be said
to be fragile around that very value. For convenience, we  look for sign changes
in $\partial \log F/\partial T$ instead of $\partial  F/\partial T$. It can be seen
that
 \be
 \frac{\partial \log F}{\partial T}= \frac{H^3(T,p)}{\partial H(T,p)/\partial T}
 \left[\frac{9}{2}-\frac{({\partial H(T,p)}/{\partial T})^2}{H^3(T,p)}
 \right].
 \ee
Therefore, the fragility condition is
\be
\frac{H'^2}{H^4}=\frac{9}{2}\label{fragility},
\ee
which, as will be shown immediately,  is all we need to answer the question
of whether inflation is rigid in tachyon cosmology. 

In many inflationary models, there is a long period during which the field  slowly rolls  down the potential, which means the kinetic energy is negligible as compared to the potential one. That will be the case if
\begin{equation}H'^2\ll \frac{9}{4} H^4.
\end{equation}
From here, rigidity of inflation in the slow-roll regime follows automatically.

Nevertheless,  kinetic energy takes over eventually, and slow-roll inflation ceases, so we must address the question of whether inflation is rigid in general.
The condition for inflation to occur is $\rho+3p<0$ (violation of the strong energy condition), which alternatively reads $\dot T^2<2/3$. Expressed
 in a yet more convenient it becomes $H'^2/H^4<3/2$, which is far from (\ref{fragility}), so that one concludes inflation is rigid.

Summarizing, we have proved tachyonic inflation is rigid using a procedure that allows spotting changes in the attractor solutions just by checking at the value of some function of a simple function of the Hubble factor and its derivative with respect to the tachyon field. A possible extension of this work is the study of rigidity in k-inflation \cite{kinflation}, considering, for instance recently found uniparametric family of  models \cite{Chimento} which include
those studied here as particular examples.
\section*{Acknoledgements}
We thank L.P. Chimento for conversations, and acknowledge support from the University of the Basque Country through research grant 
UPV00172.310-14456/2002. JMA also acknowledges support from the Spanish Ministry of Science and Technology through research grant  BFM2000-0018.
 RL also acknowledges support from the Basque Government through fellowship BFI01.412, and from the Spanish Ministry of Science and Technology
jointly with FEDER funds through research grant  BFM2001-0988. 
 

\begin{thebibliography}{99} 
 \bibitem{WMAP}C. L. Bennett et al., Astrophys. J. Suppl. {\bf 148}, 1 (2003);
 E. Komatsu et al.,Astrophys. J. Suppl. {\bf 148}, 119 (2003);
G. Hinshaw et al., Astrophys. J. Suppl. {\bf 148}, 135 (2003);
 D. N. Spergel et al., Astrophys. J. Suppl. {\bf 148}, 175 (2003);
 H.V. Peiris et al., Astrophys. J. Suppl. {\bf 148}, 213 (2003).
\bibitem{sen}A. Mazumdar, S. Panda and A. P\'erez-Lorenzana, Nucl. Phys. B {\bf 614}, 101 (2001);
 A. Sen,  JHEP {\bf 0204}, 048 (2002);  JHEP, {\bf 0207}, 065 (2002).
\bibitem{others}G.W. Gibbons, Phys.Lett. B {\bf 537},  1 (2002);
M. Fairbarn and M.H.G. Tytgat, Phys. Lett. B {\bf 546}, 1  (2002);T. Padmanabhan, Phys. Rev. D {\bf 66}, 021301 (2002);T. Padmanabhan and T. Roy Choudhury, Phys. Rev. D {\bf 66}, 081301 (R) (2002);   M. Sami, P. Chingangbam, and T. Qureshi, Phys. Rev. D {\bf 66}, 043530 (2002);  A. Feinstein, Phys. Rev. D {\bf 66}, 063511 (2002); J.-c. Hwang and H. Noh, Phys. Rev. D {\bf 66}, 084009 (2002);G. Shiu and I. Wasserman Phys.Lett. B {\bf 541}, 6 (2002); J.-g. Hao and X.-z. Li, Phys. Rev. D {\bf 66}, 087301 (2002);Y.-S. Piao, R.-G. Cai, X. Zhang, and Y.-Z. Zhang, Phys. Rev. D {\bf 66}, 121301 (2002); 
 B. Chen, M. Li, F.-L. Lin, JHEP {\bf 0211},  050 (2002); J.S. Bagla , H.K. Jassal, and T. Padmanabhan, Phys. Rev. D67, 063504 (2003);
X.-z. Li and X.-h. Zhai, Phys. Rev. D {\bf 67}, 067501 (2003), D.A. Steer  and F.Vernizzi, hep-th/0310139; G.W. Gibbons, Class. Quant. Grav. {\bf 20}, S321 (2003); V. Gorini, A. Yu. Kamenshchik, U. Moschella, V. Pasquier, hep-th/0311111.
\bibitem{attractors} Z.-K. Guo, Y.-S. Piao, R.-G. Cai, Y.-Z. Zhang, Phys. Rev. D {\bf 68}, 043508 (2003); L.R.W. Abramo and F. Finelli, Phys. Lett. B {\bf 575}, 165 (2003).


\bibitem{brane}S. Mukohyama,
Phys. Rev. D {\bf 66}, 024009 (2002); M.C. Bento, O. Bertolami, and A. A. Sen, Phys. Rev. D {\bf 67}, 063511 (2003);
M.C. Bento, N.M.C. Santos, A.A. Sen, astro-ph/0307292.


\bibitem{critics}
A. Frolov, L. Kofman, A. Starobinsky, 
Phys. Lett. B {\bf 545},  8 (2002); L. Kofman, A.Linde,
 JHEP {\bf 0207},  004 (2002). 
\bibitem{stst} A.A. Andronov and L.S. Pontryagin, Dokl. Akad. Nauk. SSSR {\bf 14}, 247 (1937).
\bibitem{coley}A.A. Coley and R.K. Tavakol, Gen. Rel. Grav. {\bf 24}, 835 (1992).
\bibitem{tavakol}R.K. Tavakol and G.F.R. Ellis, Phys. Lett. A {\bf 130}, 217 (1988);
L. Farina-Busto and R.K. Tavakol, Europhys. Lett. {\bf 11}, 493 (1990);
R.K. Tavakol, Brit. J. Phil. Sci. {\bf 42}, 147 (1991).
\bibitem{lidsey}J.E. Lidsey, Gen. Rel. Grav. {\bf 25}, 399 (1993).
\bibitem{kinflation} C. Armend\'ariz-Pic\'on, T. Damour and V. Mukhanov, Phys. Lett. B {\bf 458}, 209 (1999).
\bibitem{Chimento} L.P. Chimento, astro-ph/0311613.
 \end{thebibliography}
\end{document}